\documentclass[prb,twocolumn]{revtex4}
\usepackage{amsmath}
\usepackage{graphicx}

\newcommand{\mysig}{\mathcal{S}}
\newcommand{\expe}[1]{\mathrm{e}^{#1}}
\newcommand{\ddd}{\mathrm{d}}
\newcommand{\bfvec}[1]{\mathbf{#1}}

\begin{document}

\title{Dissipative tunneling and orthogonality catastrophe in molecular transistors}
\author{Stephan Braig$^1$ and Karsten Flensberg$^{1,2}$}
\affiliation{$^1$Laboratory of Atomic and Solid State Physics,
Cornell University, Ithaca, NY 14853\\ $^2$\O rsted Laboratory,
Niels Bohr Institute fAPG, Universitetsparken 5, 2100 Copenhagen,
Denmark.}
\date{\today}

\begin{abstract}
Transport through molecular devices with weak tunnel coupling to the leads but with strong coupling to a single vibrational mode is considered in the case where the vibration is damped by coupling to the environment. In particular, we investigate what influence the electrostatic coupling of the charge on the molecule to the vibrational modes of the environment has on the $I$-$V$ characteristics. We find that, for comparable characteristic length scales of the van-der-Waals and electrostatic interaction of the molecule with the environmental vibrational modes, the $I$-$V$ characteristics are qualitatively changed from what one would expect from orthogonality catastrophe and develop a steplike discontinuity at the onset of conduction. For the case of very different length scales, we recover dissipation consistent with modeling the electrostatic forces as an external influence on the system comprised of molecule and substrate, which implies the appearance of orthogonality catastrophe, in accord with previous results.
\end{abstract}
\maketitle

\section{Introduction}
\label{sec:intro}
The possibility of creating devices on a molecular level has opened up the field of single-molecule electronics in recent years. The transport properties of such mesoscopic systems have been investigated in a number of experiments.~\cite{reed97,park00,park02,lian02,smit02,zhit02,pasu03,Yu04} Of particular interest has been the influence of strong electron-phonon coupling on electron transport,~\cite{park00,park02,zhit02}  which manifests itself as emission and absorption of vibrational quanta observable in the excitation spectra. One example is the series of experiments by Park {\it et al.}~\cite{park00} where it was shown that the current through a single $C_{60}$ molecule was strongly coupled to a single vibrational mode. 

A large amount of theoretical works has dealt with the problem of tunneling through a single molecular electronic level that is coupled to phonon modes. Since in many experimental realizations the tunnel coupling of the molecule to the leads is rather weak compared to the other energy scales in the problem, transport is dominated by Coulomb-blockade effects and occurs sequentially. An approach based on rate equations can thus be justified and has been used in a number of recent papers.~\cite{boes01,mcca02,mitr03,brai03}
Physically, it is an essential question how the excited vibrational levels are allowed to relax, either through coupling to the environment, for example the phonons or plasmons of the metal substrate, or by virtue of the tunneling electrons.~\cite{aji03, flen03a,mitr03a} In the case where the relaxation of the vibrational mode is faster than the tunneling rate one can assume an equilibrium phonon distribution.

The coupling between the vibrational mode of the molecule and the environment depends strongly on which vibrational mode is considered. For intra-molecular vibrations the lifetime can be very long.~\cite{gure98,patt02} However, we were able to show in a recent paper~\cite{brai03} that the vibrational mode associated with the center of mass motion of the molecule is coupled more strongly to the environment and is thus exposed to an effective damping mechanism. The electron-phonon effects seen in the experiments of Ref.~\onlinecite{park00} are suggested to be due to such a center of mass motion.

\begin{figure}[pb]
\setlength{\unitlength}{1cm}
\begin{picture}
(8.5,4)(0,0) \put(0,0){\includegraphics[height=3.4cm]{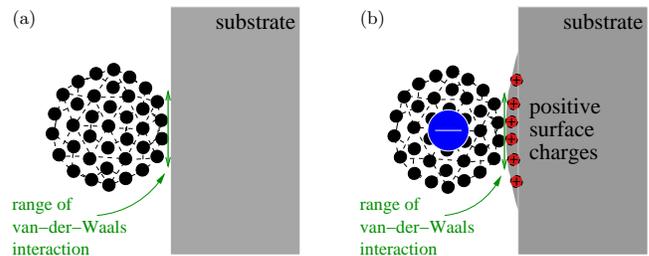}}
\end{picture}
\caption{Illustration of the system considered in this paper. {\bf (a)} The molecule is attached to a substrate, e.g. by van-der-Waals interactions. The molecule can perform center-of-mass oscillations about its equilibrium position, which also exerts forces on the substrate. We label this interaction the \emph{van-der-Waals} interaction in the following, and  we focus on the linear regime where the interaction strength is independent of the position of the molecule. {\bf (b)} When an electron hops onto the molecule, the force created by induced surface charges causes a shift of the equilibrium position of the oscillator and of the vibrational modes of the substrate. For a pointlike charge on the molecule, the shift of the equilibrium positions takes place in an effective dipole field (multipole field for non-pointlike charges). We label this interaction the \emph{electrostatic} interaction. 
} \label{fig:dipole}
\end{figure}

In the present paper, we investigate how screening of the charge on the molecule influences the damping mechanism. A sketch of the physical setup that serves as starting point is shown in Fig.~\ref{fig:dipole}.
If the molecule is not charged, it is held in place by an interaction of the van-der-Waals type. It can perform center-of-mass oscillations about its equilibrium position, and the characteristic energy $\hbar\omega$ of such an oscillation was found to be $\sim$~5meV.~\cite{park00} These oscillations also influence the substrate to which the molecule is attached, and we will call this interaction the \emph{van-der-Waals} interaction in the following. On the other hand, if the molecule is charged, there will be additional electric forces on the molecule, which are either due to static electric charges on the surface, e.g. because of impurities, or due to surface charges in the substrate that are induced by the charge on the molecule. Hence, when the electron tunnels onto the molecule, it results not only in a force on the molecule itself but also in electrostatic forces on the substrate to which it is attached, and the equilibrium positions of the vibrational modes in the substrate are shifted accordingly. In the following, we will focus on the case of surface charges only, but static impurity charges can be treated analogously and have the very same qualitative effect. We will refer to this type of interaction as the \emph{electrostatic} interaction.~\cite{footnoteMagnitudeSubcouple}

\begin{figure}[pt]
\setlength{\unitlength}{1cm}
\begin{picture}
(7,8)(0,0) \put(0,0){\includegraphics[width=7cm]{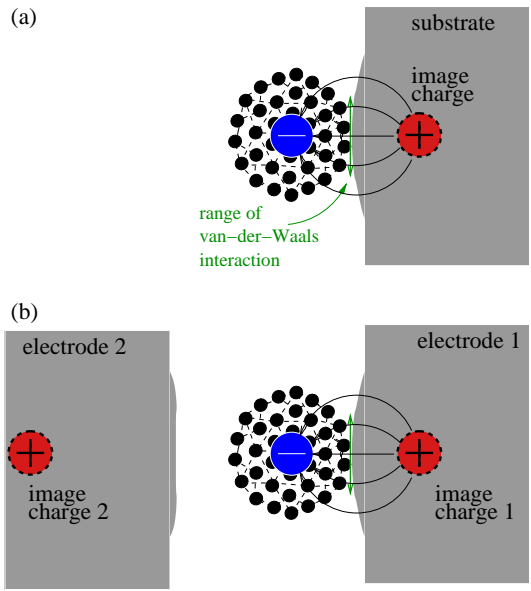}}
\end{picture}
\caption{{\bf (a)} Outside the substrate, the electric field between the charged molecule and the substrate is equivalent to the case of an image charge located inside the substrate and no surface charges. The shift of the equilibrium positions of the vibrational modes of the substrate can thus be understood to occur in an effective dipole field for pointlike charges (multipole field in general).  Since the shift of the equilibrium positions then has an integrable long-range part (see text), the small-frequency part of the phonon spectrum is not shifted, and thus there will be no orthogonality catastrophe associated with the tunneling process. The situation is different in {\bf (b)}, where the molecule interacts with image charges in both electrodes. If the two electrodes can be considered as mechanically independent, there will be long-range components of the displacement fields (the second electrode can be thought of as an external influence), and the orthogonality is restored. This case was considered in our previous analysis Ref.~\onlinecite{brai03}.
} \label{fig:Image}
\end{figure}

Under the influence of the Coulomb interaction, the surface is attracted to the molecule (and vice versa), and substrate atoms will be shifted away from their equilibrium positions until van-der-Waals forces and electrostatic forces are in equilibrium. This shift, however, occurs in an effective dipole field (or multipole field for charges that are not pointlike) since the field between molecule and substrate due to the additional charge on the molecule is equivalent to the case without surface charges but an image charge located inside the substrate, cmp. Fig.~\ref{fig:Image}. Only in the limit of a very short ranged van-der-Waals interaction compared to the separation of charge and image charge could we approximate the electric field at the surface as constant, and thus monopolar. Such a mismatch of length scales is rather unlikely for a situation as depicted in Figs.~\ref{fig:dipole} and \ref{fig:Image}(a). However, a different experimental geometry as in Fig.~\ref{fig:Image}(b) can introduce monopolar effects.~\cite{brai03}

The shift of the equilibrium positions $\bfvec{a}_\bfvec{k}$ for each phonon mode $\bfvec{k}$ can be calculated by resorting to the  dynamical matrix $\bfvec{D}(\bfvec{k},\bfvec{k}')$ which appears in the harmonic expansion of the Hamiltonian (see also Ref.~\onlinecite{loui95}):
\begin{equation}
H_\mathrm{harm}=\frac12 \sum_{\bfvec{k}\,\bfvec{k}'}\bfvec{a}_\bfvec{k} \bfvec{D}(\bfvec{k},\bfvec{k}')\bfvec{a}_{\bfvec{k}'}
\end{equation}
If we denote the electrostatic force on the substrate as $\bfvec{G}(\bfvec{r})$ and choose a phonon eigenmode basis such that $\bfvec{D}(\bfvec{k},\bfvec{k}')$ is diagonal, we can see that force equilibrium demands
\begin{equation}
\bfvec{D}(\bfvec{k})\bfvec{a}_\bfvec{k}\sim\sum_\bfvec{r}\bfvec{G}(\bfvec{r})\expe{i\bfvec{k}\bfvec{r}}
\end{equation}
where the right hand side is the Fourier component of the force $\bfvec{G}$ acting on the plane wave phonon state $\bfvec{k}$. The long-wavelength behavior can give rise to a version of orthogonality catastrophe, i.e. zero overlap between the ground states corresponding to the shifted and unshifted equilibrium positions. For small wavevectors $k$, we know that $\bfvec{D}(k)\sim k^2$, see Ref.~\onlinecite{ashc76}. Expanding $\bfvec{G}$ in a multipole expansion, and expanding the exponential for small $k$, we can see that $\mathrm{a}_\bfvec{k}\sim k^{-1}$ for a dipole force since the volume integral of any multipole field is zero by definition,~\cite{jack99} whereas $\mathrm{a}_\bfvec{k}\sim k^{-2}$ for a monopole field since this corresponds to the constant part in the multipole expansion. This difference in long-wavelength behavior gives rise to nonzero (dipole) or zero (monopole) overlap between the ground states in three dimensions, as one can see from the following simple argument. Let $\eta$ be the ground-state overlap between a set of harmonic quantum oscillators and its shifted counterparts, then
\begin{align}
\eta\sim\exp\left(-\sum_\bfvec{k}\bfvec{a}_\bfvec{k}^2/2\ell^2_\bfvec{k}\right)\label{phononeta}
%\eta\sim\exp\left(-\sum_\bfvec{k} \bfvec{x}_\bfvec{k}^2/2\ell^2_\bfvec{k}-\sum_\bfvec{k} (\bfvec{x}_\bfvec{k}-\bfvec{a}_\bfvec{k})^2/2\ell^2_\bfvec{k}\right)
\end{align}
where $\ell_\bfvec{k}^2=\hbar/m_\bfvec{k}\omega_\bfvec{k}$ is the oscillator quantum with $\ell^2_\bfvec{k}\sim k^{-1}$ for small $k$. Converting the sum to an integration, we see that the question of zero or nonzero overlap is decided by both dimensionality and the small-$k$ behavior of the equilibrium shifts $\bfvec{a}_\bfvec{k}$. In particular, 
\begin{align}
\eta\sim\exp\left(-\int\ddd k\,k^{d} \bfvec{a}_\bfvec{k}^2/A\right),
\end{align}
where $d$ is the dimensionality, and $A$ is a normalization factor. Thus, in one dimension, we encounter orthogonality catastrophe for both monopolar and dipolar shifts, whereas in two and three dimensions we see that dipolar equilibrium shifts feature nonzero overlap!

\begin{figure}[pt]
\setlength{\unitlength}{1cm}
\begin{picture}
(6,4.5)(0,0) \put(-1,0){\includegraphics[width=7.5cm]{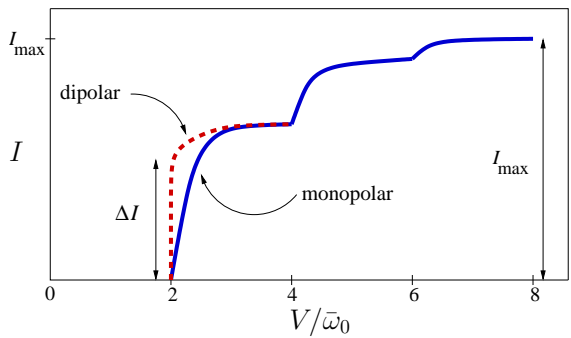}}
\end{picture}
\caption{Sketch of the difference between dipolar (dashed line) and monopolar case (solid line) in the $I$-$V$ characteristics. We depict the situation for $\varepsilon_0$=2$\bar{\omega}_0$ and $V_L=V/2=-V_R$ in the notation of Secs.~\ref{sec:model}, \ref{sec:rate}, and \ref{sec:with}. The dipolar curve features a finite step of size $\Delta I$ at the onset of conduction, whereas the monopolar curve rises according to a power law.
} \label{fig:IVEx}
\end{figure}

It is this effect that we investigate in the present paper in the context of how a single vibrational mode of the molecule dissipates energy into a phononic environment. Again referring to Fig.~\ref{fig:Image}, if the range of the van-der-Waals interaction $D$ is comparable with the dipole length scale (set by the separation of charge and image charge), we expect nonzero overlap between the two ground states in accord with a dipolar equilibrium shift, and we should thus see a step in the $I$-$V$ characteristics, cmp. Fig.~\ref{fig:IVEx}. If, on the other hand, the van-der-Waals range is very small compared to the dipole length scale, we expect an orthogonality catastrophe corresponding to the approximately constant (monopolar) force. The latter can also occur for particular geometries, cmp. Fig.~\ref{fig:Image}(b). The current then behaves according to  a power law at the onset of conduction.~\cite{brai03} 

Another way of looking at the same problem is in terms of translational invariance. Suppose that the van-der-Waals range $D$ and the dipole length scale are comparable to each other. In this case, we could think of drawing a box around the whole system which contains both the van-der-Waals and the electrostatic interaction. Hence, all forces are internal, and we have a translationally invariant Hamiltonian. Small $D$, on the other hand, corresponds to the case that, from the point of view of the van-der-Waals interaction, the electrostatic interaction is extremely extended and spread out. Therefore, the electrostatic force acts approximately as an \emph{external} force as far as the van-der-Waals interaction is concerned and translational invariance is broken. This case can be recovered in a physical setup where the molecule is located asymmetrically in between two substrates, cmp. Fig.~\ref{fig:Image}(b). Then, image charges in the substrate farther away can approximately have the effect of an external force on the substrate closer to the molecule.

The paper is organized as follows. The model Hamiltonian is defined in Sec.~\ref{sec:model}, and in Sec.~\ref{sec:rate} we recapitulate the expression for the current derived on the basis of rate equations in Ref.~\onlinecite{brai03}. The function that describes the tunneling density of states is then solved in the case of equal spread of van-der-Waals and electrostatic interaction in the presence of the dissipative environment in Sec.~\ref{sec:with}. Finally, a discussion and summary of the results can be found in Sec.~\ref{sec:summary}.

\section{Model Hamiltonian}

\label{sec:model}

We consider a model of one single spin-degenerate molecular level
coupled to two leads. The single level is coupled to the vibrational
mode of the molecule through the charge on the molecule, as are the vibrational modes of the substrate. The coupling between the oscillator and the environment is included as a linear coupling to a bath of harmonic oscillators. Similar to the model studied in our previous paper,~\cite{brai03} the model Hamiltonian then reads
\begin{align}
H&=H_{LR}^{{}}+H_{D}^{{}}+H_{B}^{{}}+H_{\mathrm{bath}}^{{}}+H_{DB}\nonumber\\
&\qquad +H_{D\mathrm{bath}}+H_{B\mathrm{bath}}^{{}}+H_{T}^{{}}, \label{Hstart}
\end{align}
with
\begin{subequations}
\label{H}
\begin{align}
H_{LR}^{{}}  &
=\sum_{k\sigma,\,\alpha=L,R}\xi_{k\alpha}^{{}}c_{k\sigma,\alpha}^{\dagger
}c_{k\sigma,\alpha}^{{\vphantom{\dagger}}}\\
H_{D}^{{}}  &  =\sum_{\sigma}\xi_{0}^{{}}d_{\sigma}^{\dagger}d_{\sigma}^{{\vphantom{\dagger}}
}+Un_{d\uparrow}n_{d\downarrow},\\
H_{\mathrm{B}}^{{}}  &  =\frac{p_{0}^{2}}{2m_{0}^{{}}}+\frac{1}{2}m_{0}^{{}}\omega_{0}^{2}x_{0}^{2},\\
H_{DB}^{{}}  &  =\lambda x_{0}^{{}}\sum_{\sigma}d_{\sigma}^{\dagger}d_{\sigma}^{{}},\\
H_{\mathrm{bath}}^{{}}  &  =\sum_{j}\left(  \frac{p_{j}^{2}}{2m_{j}}+\frac{1}{2}m_{j}\omega_{j}^{2}x_{j}^{2}\right)  ,\\
H_{B\mathrm{bath}}^{{}}  &
=\sum_{j}\beta_{j}^{{}}x_{j}^{{}}x_{0}^{{}}, \label{HBbath}\\
H_{T}^{{}} &=\sum_{k\sigma,\,\alpha=L,R}t_{k\alpha}^{{}}c_{k\sigma,\alpha}^{\dagger
}d_{\sigma}^{{}}+h.c.
\end{align}
where the $c_{k\sigma,\alpha}^{\dagger},\, c_{k\sigma,\alpha}^{{\vphantom{\dagger}}}$ and the $d_{\sigma}^{\dagger},\,d_{\sigma}^{{\vphantom{\dagger}}}$ are creation and annihilation operators for the leads and the molecule, respectively, $x_{0}^{{}}$ is the oscillator degree of freedom, 
$\{x_{j}\}$ describes the set of environmental degrees of freedom
and $m_{j}$, $\omega_{j}$ their respective masses and frequencies,
$\xi_{0}$ is the onsite energy, and $U$ is comprised of the Coulomb interaction
and the energy cost for having double occupancy on the molecule. The coupling
constants for the electron-oscillator interaction and the oscillator-bath interaction are
$\lambda$ and $\{\beta_{j}\},$ respectively. The lead electron
energies are given by
\end{subequations}
\begin{equation}
\xi_{k\alpha}=\varepsilon_{k\alpha}-\mu_{\alpha},
\end{equation}
where $\mu_{\alpha}^{{}}$ is the chemical potential of lead $\alpha$. 
The tunneling amplitudes $t_{k\alpha}$ in the tunneling term $H_T$ could in principle also depend on the oscillator position, which should be a small effect for the experimental realizations in Refs.~\onlinecite{park00} and~\onlinecite{park02}.~\cite{brai03}

We can explicitly write down how the $\beta_j$ are connected to the displacement $u(\bfvec{r})$ perpendicular to the substrate if we assume that the force acting between the molecular vibrational mode and the substrate phonons is spread over the substrate according to a normalized distribution $f(\bfvec{r})$, i.e. $\int \ddd\bfvec{r}f(\bfvec{r})=1$. If we model the interaction as a harmonic spring potential, we can write
\begin{equation}
H_{\mathrm{int}}=\frac12 m_0\omega_0^2\left[x_0-\int\ddd\bfvec{r}f(\bfvec{r})u(\bfvec{r})\right]^2.
\end{equation}
Expanding the square, we can combine the term $\frac12m_0\omega_0^2\left[\int\ddd\bfvec{r}f(\bfvec{r})u(\bfvec{r})\right]^2$ with the part of the Hamiltonian that describes the ``free'' environmental bath and diagonalize the combination by introducing new coordinates $x_j$. These are related to $u(\bfvec{r})$ by a unitary transformation as 
\begin{equation}
u(\bfvec{r})=\sum_j c_j(\bfvec{r})x_j, \label{ujxj}
\end{equation}
 with appropriately chosen coefficient functions $c_j(\bfvec{r})$. Defining
\begin{equation}
\beta_j\equiv -m_0\omega_0^2\int \ddd\bfvec{r}f(\bfvec{r})c_j(\bfvec{r}),\label{betajcj}
\end{equation}
we immediately see that we recover the structure of our original Hamiltonian $H$. The advantage of the form of $H_\mathrm{int}$ is that  its translational invariance with respect to shifting both $x_0$ and $u(\bfvec{r})$ by the same constant displacement is immediately apparent since the distribution function $f(\bfvec{r})$ is normalized to 1.

The difference to Ref.~\onlinecite{brai03} is the newly included coupling term $H_{D\mathrm{bath}}$ between the charge on the molecule  and the bath of vibrational modes,
\begin{eqnarray}
H_{D\mathrm{bath}}=\sum_j \lambda_j x_j \sum_{\sigma}d_{\sigma}^{\dagger}d_{\sigma}^{\vphantom{dagger}}.
\end{eqnarray}
It is again instructive to rewrite this term as a function of $u(\bfvec{r})$ and a normalized distribution function $g(\bfvec{r})$ that characterizes the form and range of the electrostatic interaction between molecule and substrate:
\begin{eqnarray}
H_{D\mathrm{bath}}=-\lambda \int \ddd\bfvec{r} g(\bfvec{r}) u(\bfvec{r})\sum_{\sigma}d_{\sigma}^{\dagger}d_{\sigma}^{\vphantom{dagger}}, \label{HDbathint}
\end{eqnarray}
so that the electrostatic force on the substrate $\bfvec{G}(\bfvec{r})$  mentioned in Sec.~\ref{sec:intro} has the form $\bfvec{G}(\bfvec{r})=\lambda g(\bfvec{r})\sum_{\sigma}d_{\sigma}^{\dagger}d_{\sigma}^{\vphantom{dagger}}$.
Since $u(\bfvec{r})$ can be written as a linear superposition of the eigenmodes $x_j$, see Eq.~(\ref{ujxj}), we can express the coupling constants $\lambda_j$ in terms of the distribution function $g(\bfvec{r})$ as
\begin{eqnarray}
\lambda_j = -\lambda\int \ddd\bfvec{r} g(\bfvec{r}) c_j(\bfvec{r})\label{lambdajcj}.
\end{eqnarray}
Given Eq.~(\ref{HDbathint}), translational invariance of the total Hamiltonian with respect to shifting both $x_0$ and $u(\bfvec{r})$ by some constant displacement vector is ensured automatically if $g(\bfvec{r})$ is normalized properly, i.e. $\int\ddd\bfvec{r} g(\bfvec{r})=1$.

We now want to relate the coupling constants in the boson-bath
coupling to the finite damping of the vibrational mode, which can
be accomplished by studying the classical equations of motion.
After removing the bath degrees of freedom, we end up with the
following equation of motion in the frequency domain, neglecting the coupling terms $H_{DB}^{{}}$ and $H_{D\mathrm{bath}}$ which will be removed by a unitary transformation below:
\begin{eqnarray}
[\omega^{2}-\omega_0^{2}-\mysig(\omega)]x_{0}(\omega)=0,\label{xeqmfreq}
\end{eqnarray}
where we have defined
\begin{eqnarray}
\mysig(\omega)=\frac1{m_0}\sum_j\frac{\beta_j^2}{m_j}
\frac{1}{(\omega+i\eta)^2-\omega_j^2}, \label {mysigdef}
\end{eqnarray}
which is complex in general and gives rise to frictional damping
and a frequency shift of the bare frequency $\omega_0$.
In Ref. \onlinecite{brai03}, we explicitly calculated
$\mysig(\omega)$ for the case of a molecule attached to a
semi-infinite substrate. We will use these results throughout this paper, again assuming that the van-der-Waals force $\mathcal{F}$ exerted by the molecule, cmp. Fig.~\ref{fig:dipole}(a), is directed perpendicular to the substrate surface and cylindrically symmetric. Also, we assume that both the van-der-Waals and the electrostatic forces only couple to the substrate surface, i.e., $g(\bfvec{r}),f(\bfvec{r})\sim \delta(x)$, where the $x$ axis is perpendicular to the substrate surface. The van-der-Waals force $\mathcal{F}$ on the molecule is thus understood to be of the form
\begin{eqnarray}
\mathcal{F}=-m_0\omega_0^2\left[x_0-\int_0^\infty\!\!\!  2\pi r f(r) u_x(r)dr\right], \label{totalforce}
\end{eqnarray}
where $u_x(r)$ is the perpendicular displacement on the surface of the substrate, and $r$ is the radial direction in the surface plane. For explicit numerical evaluations later in this paper, we will use $f(r)=\expe{-r/D}/2\pi D^2$ just as in our previous paper, with $D$ on the order of the width $D_0$ of the molecule.~\cite{brai03}

We now eliminate the coupling terms $H_{DB}^{{}}$ and $H_{D\mathrm{bath}}$ of the Hamiltonian (\ref{Hstart}) by a unitary transformation similar to the
one used in the independent boson model,~\cite{mahan} at the cost
of introducing displacement operators in the tunneling term.
However, since we have a somewhat more complicated system here due
to the coupling to the bosonic bath, the unitary transformation in
Ref.~\onlinecite{mahan} has to be generalized. We define the
transformation
\begin{equation}
\tilde{H}=SHS^{\dagger},\quad S=e^{-iAn_{d}^{{}}},\quad
A=p_{0}^{{}}\ell +\sum_{j}p_{j}^{{}}\ell_{j}^{{}}, \label{Sdef}
\end{equation}
where $n_{d}^{{}}=\sum_{\sigma}d_{\sigma}^{\dagger}d_{\sigma}$. Using that
\begin{equation}
\tilde{x}_{0}^{{}}=x_{0}^{{}}-\ell
n_{d}^{{}},\quad\tilde{x}_{j}^{{}}
=x_{j}^{{}}-\ell_{j}^{{}}n_{d}^{{}},
\end{equation}
both linear coupling terms, $H_{DB}^{{}}$ and $H_{D\mathrm{bath}}$, cancel if we set
\begin{eqnarray}
\ell=\frac{\lambda-\sum_j \frac{\beta_j \lambda_j}{m_j \omega_j^2} }{m_{0}^{{}}[\omega_{0}^{2}+\mysig(0)]},\qquad
\ell_{j}=\frac{\lambda_j}{m_j\omega_j^2}-\frac{\ell\beta_j}{m_j\omega_j^2}
\label{Sdef2}
\end{eqnarray}
The Hamiltonian then transforms into
\begin{equation}
\tilde{H}=H_{LR}^{{}}+\tilde{H}_{D}^{{}}+H_{B}^{{}}+H_{\mathrm{bath}}^{{}
}+H_{B\mathrm{bath}}^{{} }+\tilde{H}_{T}^{{}}, \label{Huni}
\end{equation}
where
\begin{equation}
\tilde{H}
_{T}=\sum_{k\sigma,\,\alpha=L,R}t_{k\sigma,\alpha}^{{}}\left(
c_{k\sigma,\alpha}^{\dagger}e^{iA}d_\sigma^{{}}+d_{{\sigma}}^{\dagger}
e^{-iA}c_{k\sigma,\alpha}^{{}}\right)
\end{equation}
and
\begin{equation}
\tilde{H}_{D}^{{}}=\varepsilon_{0}^{{}}\sum_\sigma
d_{{\sigma}}^{\dagger}d_\sigma^{{}}
+\tilde{U}n_{d\downarrow}^{{}}n_{d\uparrow}^{{}},
\quad\varepsilon_{0}^{{}}=\xi_{0}^{{}}-\frac{1}{2}\lambda\ell.
\end{equation}
Here we have defined
\begin{align}
\tilde{U}&\equiv U-\frac{\kappa}{m_{0}[\omega_{0}^{2}+\mysig(0)]},\,\,
\varepsilon_{0}^{{}} \equiv\xi_{0}-\frac{1}{2}\frac{\kappa}{m_{0}[\omega_{0}^{2}+\mysig(0)]},
\end{align}
where
\begin{align}
\kappa\equiv&\frac{1}{2}\lambda^2+\frac{1}{2}\sum_j\frac{\lambda_j^2}{m_j\omega_j^2}\left(1-\sum_{i\neq j}\frac{\beta_i^2}{m_i\omega_i^2}\right)\nonumber\\
&\,\,-\sum_j\frac{\beta_j\lambda_j}{m_j \omega_j^2}\left(\lambda-\sum_{i>j}\frac{\beta_i\lambda_i}{m_i\omega_i^2}\right)
\end{align}
Thus, the Coulomb repulsion $U$ is modified by both the phonon mediated interaction and the electrostatic interaction between molecule and substrate. 

It is important to note that for the particular choice $f(\bfvec{r})=g(\bfvec{r})$, i.e. when the distribution functions of electrostatic and van-der-Waals interaction are equal, we obtain from  Eqs. (\ref{betajcj}) and (\ref{lambdajcj}):
\begin{eqnarray}
\lambda_j=\frac{\lambda\beta_j}{m_0\omega_0^2}\label{specialEllj},\qquad
\ell=\frac{\lambda}{m_0\omega_0^2},\qquad \ell_j=0. 
\end{eqnarray}
Furthermore, we can quantify the deviation of $f(\bfvec{r})$ from $g(\bfvec{r})$ by defining
\begin{equation}
\epsilon(\bfvec{r})\equiv g(\bfvec{r})-f(\bfvec{r}),\,\,\,\,\label{epsdef}\epsilon_j\equiv -m_0\omega_0^2\int\ddd\bfvec{r}\epsilon(\bfvec{r})c_j(\bfvec{r}),
\end{equation}
where $\int\ddd\bfvec{r}\epsilon(\bfvec{r})=0$.
We can thus rewrite $\lambda_j$ as
\begin{equation}
\lambda_j=\frac{\lambda}{m_0\omega_0^2}\left(\beta_j+\epsilon_j\right).
\end{equation}
This immediately leads to
\begin{eqnarray}
\ell=\frac{\lambda}{m_0\omega_0^2}\left[1-\Delta(0)\right],\,\, \ell_j=\frac{\ell}{m_j\omega_j^2}\frac{\epsilon_j+\Delta(0)\beta_j}{1-\Delta(0)},\label{ellepsilon}
\end{eqnarray}
where we defined
\begin{eqnarray}
\Delta(\omega)\equiv-\frac1{m_0\bar{\omega}_0^2}\sum_n\frac{\beta_n\epsilon_n}{m_n}\frac1{(\omega+i\eta)^2-\omega_n^2}.
\end{eqnarray}
If the two distribution functions $f(\bfvec{r})$ and $g(\bfvec{r})$ deviate only slightly from each other, $\epsilon_j$ is small compared to $\beta_j$, and we see that $\Delta\ll 1$. Also note that the qualitative frequency dependence of $\Delta(\omega)$ is exactly the same as of $\mysig(\omega)$.

\section{Current formula}

\label{sec:rate} As mentioned in the Introduction, the most important
assumption here is that the tunneling rate is much smaller than
all other time scales, which means that we can assume the
vibrational degrees of freedom and the Fermi seas in the two
electrodes to be in equilibrium at all times. For simplicity, we
consider only two charge states and therefore let $U=\infty$,
which leaves us with only three states playing a role in the rate equations: empty, and occupied by
either spin up or down. Treating the tunneling term $H_T$ as a perturbation and accordingly employing Fermi's Golden Rule for the tunneling rates, the current $I$ through the molecule is given by 
\begin{equation}
I=2e\frac{\Gamma_{10}^{R}\Gamma_{01}^{L}-\Gamma
_{01}^{R}\Gamma_{10}^{L}}{(\Gamma_{01}^{{L}}+\Gamma_{01}^{{R}})+2(\Gamma_{10}^{{L}}+\Gamma_{10}^{{R}})}.\label{I}
\end{equation}
Here, we defined $\Gamma^\alpha_{ij}$ as the tunneling rate for tunneling from the state with occupation $i$ to the state occupation $j$ through lead $\alpha$ (see, for example, Ref.~\onlinecite{devogirv}):
\begin{subequations}
\label{Gamman}
\begin{align}
\Gamma_{10}^{\alpha}  &
=\Gamma_{\alpha}^{{}}\int\frac{d\omega}{2\pi}
F(\omega)n_{\alpha}^{{}}(\varepsilon_{0}+\omega),\label{Gamma10}\\
\Gamma_{01}^{\alpha}  &
=\Gamma_{\alpha}^{{}}\int\frac{d\omega}{2\pi
}F(-\omega)(1-n_{\alpha}^{{}}(\varepsilon_{0}+\omega)),
\label{Gamma01}
\end{align}
with 
\end{subequations}
\begin{equation}
F(\omega)   =\int_{-\infty}^{\infty}dt\,F(t), \,\,\, F(t)=\langle e^{iA(t)}e^{-iA}\rangle,\label{Ftdef}
\end{equation}
where the operator $A$ is defined in Eq.~(\ref{Sdef}). Furthermore, we work in the wide-band limit with bare tunneling rates $\Gamma_{\alpha}^{{}}$=$2\pi\sum_{k}|t_{k\alpha}|^{2}\delta(\xi_{k})$, and we intoduced the Fermi distributions of the two leads $n_{\alpha}(\varepsilon)$=$(e^{\beta(\varepsilon-eV_{\alpha})}+1)^{-1}$.

\section{Influence of the coupling to the environment}

\label{sec:with}

In presence of coupling to the environment, the evaluation of the
function $F(t)$ in Eq.~(\ref{Ftdef}) is in principle
straightforward since the Hamiltonian is quadratic in the oscillator and
bath degrees of freedom. We obtain
\begin{equation}
F(t)=\exp\left(  B(t)-B(0)\right)  ,\quad B(t)=\langle
A(t)A(0)\rangle_{0}, \label{Bt}
\end{equation}
where the operator $A$ is defined in
Eqs.~(\ref{Sdef}) and (\ref{Sdef2}). The expectation value
$\langle\dots\rangle_{0}^{{}}$ is to be evaluated with respect to
$\tilde{H}$ without the tunneling term. We follow the same procedure as in our previous paper~\cite{brai03} and use the fluctuation-dissipation theorem,
\begin{equation}
B(\omega)=-2\operatorname{Im}[B^{R}(\omega)](1+n_{B}^{{}}(\omega)),
\label{flucdis}
\end{equation}
to express $B(t)$ in terms of the corresponding retarded Green's function
\begin{equation}
B^{R}(t)=-i\theta(t)\langle\lbrack A(t),A(0)]\rangle_{0}.
\end{equation}
Here, $n_{B}(\omega)=(e^{\beta\omega}-1)^{-1}$ is the usual Bose
function. We obtain
\begin{align}
B^R(\omega) =& \, \ell^2 m_0\left(\frac{\omega^2}{\omega^2-\bar{\omega}_0^2-\bar{\mysig}(\omega)}-1\right) \nonumber\\
&+2\frac{\ell^2m_0}{1-\Delta(0)}\frac{\Delta(0)\bar{\mysig}(\omega)-\bar{\omega}_0^2\bar{\Delta}(\omega)}{\omega^2-\bar{\omega}_0^2-\bar{\mysig}(\omega)}\nonumber\\
&+\frac{\ell^2 m_0}{[1-\Delta(0)]^2}\left[\Delta(0)^2\bar{\mysig}(\omega)-2\Delta(0)\bar{\Delta}(\omega)\right]\nonumber\\
&+ \frac{\ell^2 m_0}{[1-\Delta(0)]^2} \sum_{j}\frac{\epsilon_j^2}{m_j\omega_j^2}\frac1{(\omega+i\eta)^2-\omega_j^2}\nonumber\\
&+\frac{\ell^2 m_0}{\omega^2[1-\Delta(0)]^2}\frac{[\Delta(0)\bar{\mysig}(\omega)-\bar{\omega}_0^2\bar{\Delta}(\omega)]^2}{\omega^2-\bar{\omega}_0^2-\bar{\mysig}(\omega)}\label{fullBR},
\end{align}
where we have defined $\bar{\mysig}(\omega)\equiv \mysig(\omega)-\mysig(0)$, $\bar{\Delta}(\omega)\equiv \Delta(\omega)-\Delta(0)$, and the experimentally observable renormalized frequency $\bar{\omega}_0^2\equiv \omega_0^2+\mysig(0)$.

If we choose a very wide-ranged distribution $g(\bfvec{r})$, as compared to $f(\bfvec{r})$, then the function $\epsilon(\bfvec{r})$ will be close to $-f(\bfvec{r})$, see Eq.~(\ref{epsdef}). This leads to $\epsilon_j\approx-\beta_j$, and upon insertion into Eq.~(\ref{ellepsilon}) we recover the expressions for $\ell$ and $\ell_j$ that we found previously in Ref.~\onlinecite{brai03} since then $\Delta(\omega)\approx\mysig(\omega)$. In effect, the electrostatic interaction is thus so widespread compared to the van-der-Waals interaction that it can be considered as an external influence on the system. In this case, we are justified to neglect the term $H_{D\mathrm{bath}}$ in our Hamiltonian altogether since translational invariance is broken by this external force. It is a matter of straightforward algebra to show that $B^R(\omega)$ indeed reduces to the same expression as in Ref.~\onlinecite{brai03}.

If, on the other hand, the spread of the two interactions over the surface is comparable to each other, i.e. $g(\bfvec{r})\approx f(\bfvec{r})$, we have the situation where $\epsilon_j\ll \beta_j$, and $\Delta\ll 1$ in Eq.~(\ref{ellepsilon}). 
The first term in Eq.~(\ref{fullBR}) is of zeroth order in $\epsilon_j$, the second term is of first order, and all the remaining terms are of second order. Furthermore, it is important to see that the imaginary part of all terms except of the last one are convergent for small $\omega$ since the small-frequency dependence of Im$\bar{\Delta}(\omega)$  is identical to that of Im$\bar{\mysig}(\omega)\sim\omega$. However, the last term is of second order in $\epsilon_j$, so that to leading order, Im$B^R(\omega)$ is convergent for small $\omega$ if the distribution functions $f(\bfvec{r})$ and $g(\bfvec{r})$ feature comparable ranges.

How small the $\epsilon_j$ really have to be in comparison to the $\beta_j$ (i.e., how similar in range $f(\bfvec{r})$ and $g(\bfvec{r})$ have to be) in order for such a first-order expansion to make sense can be deduced from the following argument: from our previous analysis~\cite{brai03} we know that the divergent terms in $\mathrm{Im}B^R(\omega)$ for small $\omega$ behave like $C/\omega$, where $C$ is a numerical constant of second order in the $\epsilon_j$. The divergence shows its predominant influence in the large-time behavior of $F(t)$:
\begin{equation}
F(t)=\exp\left(\int_{-\infty}^{\infty}
\frac{d\omega}{2\pi} (\expe{-i\omega t}-1)B(\omega)\right). \label{F1}
\end{equation}
For large times $t$, the integral in the exponent diverges logarithmically, where the physical cutoff for the integration is given by the smallest frequency of the phonon spectrum $\omega_\mathrm{min}\ll \bar{\omega}_0$.~\cite{footnoteOmegaminSubcouple}  In order for a first-order expansion in the $\epsilon_j$ to hold, we therefore have to demand that the following be true:
\begin{equation}
\ln\left[\frac{\bar{\omega}_0}{\omega_\mathrm{min}}\right]|C|\ll1,\,\,\,\mathrm{i.e.}\,\,\,\, \frac{\bar{\omega}_0}{\omega_\mathrm{min}}\ll \expe{1/|C|}.
\end{equation}
Even if $\epsilon_j\ll\beta_j$ is enforced only moderately, this condition is not very restrictive. We should therefore expect a step in the conductance even if $f(\bfvec{r})$ and $g(\bfvec{r})$ are more than just ``slightly'' different!

In the following, we will examine the particular case $f(\bfvec{r})=g(\bfvec{r})$, i.e. $\lambda_j=\lambda\beta_j/m_0\omega_0^2$, see Eq.~(\ref {specialEllj}), for which the small-frequency behavior changes quite dramatically. (This is the case most different from our previous analysis in Ref.~\onlinecite{brai03}.) Then we have $\ell_j=0$, and thus
\begin{align}
B^R(\omega)&=\ell^2 m_0\left(\frac{\omega^2}{\omega^2-\bar{\omega}_0^2-\bar{\mysig}(\omega)}-1\right),
\end{align}
which leads to
\begin{equation}
B(\omega)=-\frac{4g}{\bar{\omega}_0}\frac{\omega^2[1+n_{B}^{{}}(\omega)]\,\mathrm{Im}\bar{\mysig}(\omega)}{\left[\omega^2-\bar{\omega}_0^2-\mathrm{Re}\bar{\mysig}(\omega)\right]^2-[\mathrm{Im}\bar{\mysig}(\omega)]^2}.\label{BSpecial}
\end{equation}
Here, we have defined the experimental parameter
\begin{equation}
g\equiv\frac12 \ell^2m_0\bar{\omega}_0=\frac12 \frac{\lambda^2\bar{\omega}_0}{m_0\omega_0^4},\label{gdef}
\end{equation}
which is a measure of how much the equilibrium position of the molecular vibrational mode is shifted compared to the oscillator quantum $\sqrt{\hbar/m_0\bar{\omega}_0}$. Note the appearance of both the renormalized and bare frequency in Eq.~(\ref{gdef}) and thus Eq.~(\ref{BSpecial}).

One can then find $F(t)$ from Eq.~(\ref{F1}). For large times $t$, the integral in the exponent is convergent and yields a nonzero finite number since Im$\bar{\mysig}(\omega)\propto \omega$ for small frequencies in the case of coupling to a semi-infinite substrate.~\cite{brai03}
In particular, we obtain 
\begin{align}
&\lim_{t\rightarrow \infty}F(t)=\exp\left(\frac{4g}{\bar{\omega}_0}\int_{-\infty}^{\infty}
\frac{d\omega}{2\pi} \frac{\omega^2[1+n_{B}^{{}}(\omega)]\,\mathrm{Im}\bar{\mysig}(\omega)}{\mathcal{G}(\omega)}\right), \label{Ftlarge}
\end{align}
where we defined 
\begin{eqnarray}
\mathcal{G}(\omega)\equiv\left[\omega^2-\bar{\omega}_0^2-\mathrm{Re}\bar{\mysig}(\omega)\right]^2-[\mathrm{Im}\bar{\mysig}(\omega)]^2
\end{eqnarray}
for the convenience of shorter formulae. Thus, we expect a $\delta$-function at $\omega=0$, which in turn results in a finite steplike discontinuity in the $I$-$V$ curves at zero temperature according to Eqs. (\ref{I}, \ref{Gamman}). We define
\begin{equation}
f(t)=F(t)-\lim_{t\rightarrow \infty}F(t),
\end{equation}
and thus obtain at zero temperature:
\begin{align}
F(\omega)=& 2\pi\left(\lim_{t\rightarrow \infty}F(t)\right)\delta(\omega)+f(\omega)
\end{align}
with
\begin{align}
f(\omega)= &-\frac{4g}{\bar{\omega}_0}\frac{\omega^2\,\mathrm{Im}\bar{\mysig}(\omega)}{\mathcal{G}(\omega)}\left(\lim_{t\rightarrow \infty}F(t)\right)\nonumber\\
&-\frac{4g}{\omega\bar{\omega}_0}\int_0^\omega\frac{d\xi}{2\pi} f(\xi)\frac{(\omega-\xi)^3\,\mathrm{Im}\bar{\mysig}(\omega-\xi)}{\mathcal{G}(\omega-\xi)}.
\end{align}
This implies that $f(\omega)\propto \omega^3$ for small $\omega$. For small $g$, we can actually approximate $f(\omega)$ by the first term, since the other terms are of higher order in $g$.

\begin{figure}[ptb]
\setlength{\unitlength}{1cm}
\begin{picture}
(6,4.5)(0,0) \put(-.5,0){\includegraphics[width=7cm]{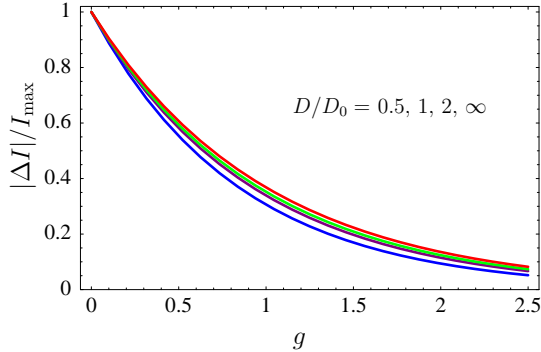}}
\end{picture}
\caption{Size of the discontinuity $|\Delta I|$ at $eV=\pm 2\varepsilon_0$ (cmp. Fig.~\ref{fig:IVEx}) calculated for a $C_{60}$ molecule on gold, with symmetric tunneling contacts $\Gamma_L=\Gamma_R$ (in units of the maximum current through the molecule $I_{\mathrm{max}}=4I_N/3$) as a function of $g$ and  the particular choices $D/D_0=$ 0.5,~1,~2,~$\infty$ (bottom to top). The large $D/D_0$ limit coincides with the case of no vibrational coupling to the environment for which we have $|\Delta I|\propto \expe{-g}$. The dependence on $D/D_0$ is rather weak but still visible. } \label{fig:discont}
\end{figure}
The resulting discontinuity $\Delta I$ in the $I$-$V$ characteristics calculated from Eq.~(\ref{I}) and measured in units of $I_N=e\Gamma_L\Gamma_R/(\Gamma_L+\Gamma_R)$ is given by
\begin{equation}
\frac{\Delta I}{I_N}=2 \,\mathrm{sign}(V) \left(\lim_{t\rightarrow \infty}F(t)\right) \frac{1+\Gamma_L/\Gamma_R}{2+\Gamma_L/\Gamma_R}.
\end{equation}
We explicitly calculate $\Delta I$ for the case of a $C_{60}$ molecule attached to a gold substrate, which is shown in Fig.~\ref{fig:discont} as a function of the experimental parameter $g$. The curves are parametrized by the spread $D/D_0$ of the vibrational force over the surface of the substrate, see Eqs.~(\ref{mysigdef}, \ref{Ftlarge}). However, the dependence of the jump $\Delta I$ on $D/D_0$ is only weak, and the size is comparable to the step size encountered in the case without coupling to the environment, for which we had  the same formula as above with $\lim_{t\rightarrow \infty}F(t)=\expe{-g}$.~\cite{brai03} The discontinuity appears at the voltage $eV=\pm 2\varepsilon_0$ for symmetric bias voltage $V_{L,R}=\pm V/2$, but is not present for $\varepsilon_0=0$ for reasons of symmetry. 

Unless suppressed for reasons of device geometry, such a discontinuity at onset of conduction should be visible experimentally for weak tunneling into and out of the molecule even when the effect of finite tunneling rates $\Gamma$  is taken into account. Following the derivations by Meir and Wingreen,~\cite{meir92} the differential conductance line shape is given by $A(\omega)\Gamma$, where $A(\omega)$ is the spectral function of the molecule. Taking the ground state phonon overlap $\eta$ from Eq.~(\ref{phononeta}), it can be shown that~\cite{flen03a}
\begin{align}
A(\omega)\Gamma\sim\frac{\Gamma^2 |\eta|^6}{\omega^2+\Gamma^2|\eta|^4/4}.
\end{align}
Thus, the height of the differential conductance step scales as $|\eta|^2$, whereas the characteristic width of the step scales as $\Gamma|\eta|^2$. For small enough $\Gamma$ compared to the width of the first Franck-Condon step in the $I$-$V$ curve, we therefore see that we recover a sharp step-like discontinuity in the $I$-$V$ curve. We can estimate $\Gamma$ by virtue of the maximum current through the molecule $I_{max}\sim e\Gamma$, which yields $\hbar\Gamma\sim 0.1\mu$eV since $I_{max}\sim$ 100pA for the experimental setup by Park \emph{et al.}~\cite{park00} The Franck-Condon steps, on the other hand, feature a source-drain voltage width on the order of mV in the $I$-$V$ curves, see Ref.~\onlinecite{park00}, so that the discontinuity in the $I$-$V$ curve will remain sharp even for finite tunneling.

\section{Summary and discussion}
\label{sec:summary}

We found that inclusion of the effects due to screening of the charge on the molecule can change the qualitative form of dissipation with which the molecular vibrational mode is faced. Such screening effects can be mediated by static surface charges on the substrate, or by surface charges that are induced in the substrate by the charge on the molecule. If the distribution functions that characterize the form of the electrostatic interaction and the van-der-Waals interaction of the molecule with the substrate are not too much different from each other, we encounter a step in the $I$-$V$ curves at the onset of conduction. Such a discontinuity should be observable experimentally in the weak tunneling limit even with smearing due to finite tunneling effects. This is a result qualitatively different from treating the electrostatic forces as an external influence on the molecule-substrate system. The step size is dependent on the experimental parameter $g$ that reflects the shift of the equilibrium position of the molecular mode compared to the oscillator quantum $m_0\bar{\omega}_0$ and features an exponential dependence on $g$ similar to what one would obtain for the case of no coupling to the environment. The exponent is weakly dependent on the spread $D/D_0$ of the van-der-Waals coupling between molecular mode and substrate modes, and we recover the decoupled limit for $D/D_0\rightarrow \infty$ as expected from our previous results in Ref.~\onlinecite{brai03}.

However, we also find that the results of  our previous analysis are not changed qualitatively (but can be changed quantitatively) if the electrostatic interaction is very long ranged compared to the van-der-Waals interaction, and we recover a power law for the current at onset of conduction. In particular, the limit $\lambda_j\rightarrow0$ taken in  Ref.~\onlinecite{brai03} is explicitly seen to coincide with treating the electrostatic interaction as an external influence on the system that breaks translational invariance. 

The data in Ref.~\onlinecite{park00} do not exhibit a step in the $I$-$V$ characteristics and thus feature agreement with the case of an external force as chosen in Ref.~\onlinecite{brai03}. 
It is quite possible, however, that in other experimental realizations, van-der-Waals interaction and electrostatic interaction are of comparable range, without the geometry of the system suppressing the effects that we found to arise due to the absence of external forces. According to our calculations, one would then indeed observe a significant step in the $I$-$V$ curve at the onset of conduction. 

\acknowledgments The authors would like to thank P.~W.~Brouwer, J.~P.~Sethna, and P.~Sharma for discussions.
The work was supported by the Cornell Center
for Materials Research under NSF Grant No.~DMR0079992, and by the Danish National Research
Council.

%\bibliography{../centralbib}

\begin{thebibliography}{25}
\bibitem[{foo({\natexlab{a}})}]{footnoteMagnitudeSubcouple}
\bibinfo{note}{In fact, the electromagnetic interaction changes with the
  displacement of the molecule $x_0$ from its equilibrium position ($\sim$
  width of the molecule $D_0$ away from the substrate). To linear order, we can
  Taylor expand, so that the strength of the position dependent force is
  smaller than the equilibrium electromagnetic interaction by a factor
  $x_0/D_0$. Compared to van-der-Waals and electromagnetic interaction, which
  have to be equal in equilibrium, the position dependence can thus be
  neglected.}
\bibitem[{foo({\natexlab{b}})}]{footnoteOmegaminSubcouple}
\bibinfo{note}{If we denote $L$ as the range over which the equilibrium
  positions of the substrate surface are displaced (i.e. the size of the
  ``box'' outside which there are no forces), then $\omega_\mathrm{min}\sim
  v_s/L$, where $v_s$ is the sound velocity.}
\end{thebibliography}
%\bibliographystyle{prsty}

\end{document}